\documentclass[aps,prd,preprint,tightenlines,superscriptaddress,showpacs]{revtex4}

\usepackage{graphicx} 

\def\myspecial#1{}                   

\newcommand{\Rdsk} {r_B}
\newcommand{\Rdbsk}{r_B}

\newcommand{\ddsk} {\delta_B}
\newcommand{\ddbsk}{\delta_B}

\newcommand{\Rdbs} {r_D}

\newcommand{\ddbs} {\delta_D}

\begin{document}

\preprint{
  \vbox{
    \hbox{   }
    \hbox{hep-ph/0409281}
    \hbox{\today}
  }
}

\title{\quad\\[0.5cm] \boldmath
  On $\phi_3$ Measurements Using $B^- \to D^* K^-$ Decays
}

\author{Alex Bondar\footnote{\tt a.e.bondar@inp.nsk.su} }
\affiliation{Budker Institute of Nuclear Physics, Novosibirsk}
\author{Tim Gershon\footnote{\tt gershon@bmail.kek.jp}}
\affiliation{High Energy Accelerator Research Organization (KEK), Tsukuba}

\myspecial{!userdict begin /bop-hook{gsave 280 50 translate 0 rotate
    /Times-Roman findfont 18 scalefont setfont
    0 0 moveto 0.70 setgray
    (\mySpecialText)
    show grestore}def end}

\begin{abstract}
  \noindent
  We point out that in decays of the type $B^- \to D^* K^-$,
  where the neutral $D^*$ meson is an admixture of $D^{*0}$ and $\bar{D}^{*0}$,
  there is an effective strong phase difference of $\pi$ between the cases that
  the $D^*$ is reconstructed as $D\pi^0$ and $D\gamma$.
  We consider the consequences for measurements of $\phi_3$ using these modes,
  some of which are profound and beneficial.
\end{abstract}


\maketitle

{\renewcommand{\thefootnote}{\fnsymbol{footnote}}}
\setcounter{footnote}{0}

Measurement of the Unitarity Triangle angle $\phi_3$~\cite{pdg_review} 
is a major challenge.
A number of methods have been proposed using $B^\mp \to D K^\mp$ decays,
including those where the neutral $D$ meson is reconstructed 
as a $CP$ eigenstate (GLW)~\cite{glw},
in a suppressed final state (ADS)~\cite{ads},
or in a self-conjugate three-body final state, 
such as $K_S \pi^+\pi^-$ (Dalitz)~\cite{dalitz}.
Each of these approaches, while theoretically clean,
has some difficulty for $\phi_3$ extraction due to intrinsic ambiguities
or model dependence;
these can be overcome by combining the results from different techniques,
and including other modes, such as $B^\mp \to D^* K^\mp$ and $B^\mp \to D K^{*\mp}$.
The first studies of these modes, and constraints on $\phi_3$,
have recently begun to appear from the $B$ factory experiments.
In this note, we consider the effect of reconstructing 
the neutral $D^*$ meson as either $D \pi^0$ or $D \gamma$,
and examine the consequences for $\phi_3$ measurements.
It is found that the ADS method is significantly enhanced,
and that $\phi_3$ can be extracted by applying this technique to 
$B^\mp \to D^* K^\mp$ decays alone.

The basic idea, for all of these approaches, 
is that the neutral $D^{(*)}$ meson produced in $B^- \to D^{(*)} K^-$ is an 
admixture of $D^{(*)0}$ (produced by a $b \to c$ transition) and 
$\bar{D}^{(*)0}$ (produced by a colour-suppressed $b \to u$ transition) states.
If the final state is chosen so that both $D^{(*)0}$ and $\bar{D}^{(*)0}$ 
contribute, the two amplitudes interfere.
The resulting observables are sensitive to $\phi_3$, 
the relative weak phase between the two $B$ decay amplitudes.

In the general case of $B^\mp \to D^{(*)} K^\mp$,
with $D^{(*)}$ decaying to a final state $f$,
we can write the decay rates for $B^-$ and $B^+$ ($\Gamma_\mp$), 
and the charge averaged rate ($\Gamma = (\Gamma_- + \Gamma_+)/2$) as 
\begin{eqnarray}
  \label{eq:gen_def}
  \Gamma_\mp  & \propto & 
  \Rdbsk^2 + \Rdbs^2 + 
  2 \Rdbsk \Rdbs \cos \left( \ddbsk + \ddbs \mp \phi_3 \right), \\
  \Gamma & \propto &
  \Rdbsk^2 + \Rdbs^2 + 
  2 \Rdbsk \Rdbs \cos \left( \ddbsk + \ddbs \right) \cos \left( \phi_3 \right),
\end{eqnarray}
where the ratio of $B$ decay amplitudes 
is usually defined to be less than one,
\begin{equation}
  \label{eq:Rdbsk_def}
  \Rdbsk = 
  \frac{
    \left| A\left( B^- \to \bar{D}^{(*)0} K^- \right) \right|
  }{
    \left| A\left( B^- \to D^{(*)0} K^-\right) \right|
  },
\end{equation}
and the ratio of $D^{(*)}$ decay amplitudes is then defined as
\begin{equation}
  \label{eq:Rdbs_def}
  \Rdbs = 
  \frac{
    \left| A\left( D^{(*)0} \to f \right) \right|
  }{
    \left| A\left(\bar{D}^{(*)0} \to f \right) \right|
  }.
\end{equation}
The strong phase differences between the $B$ and $D^{(*)}$ decay amplitudes 
are given by $\ddbsk$ and $\ddbs$, respectively.
Note that $\Rdbsk$ and $\ddbsk$ take different values for different $B$ decays;
the values for $B^- \to DK^-$ and $B^- \to D^*K^-$ are not the same.
On the other hand, the value of $\Rdbs$ depends only on the final state of
the $D$ decay, since the amplitudes for $D^{*0}$ and $\bar{D}^{*0}$ decays
to $D^{0}$ and $\bar{D}^{0}$, respectively,
via emission of either a pion or a photon, will cancel in the ratio.
For the GLW analysis, $\Rdbs = 1$ and $\ddbs$ is trivial (either zero or $\pi$),
while in the Dalitz analysis $\Rdbs$ and $\ddbs$ vary across the Dalitz plot,
and depend on the $D$ decay model used.
For the ADS analysis, the values of $\Rdbs$ and $\ddbs$ are not trivial,
and this is the main subject of our attention.

Let us consider the neutral $D^*$ meson produced in $B^-$ decay,
which we denote by $\tilde{D}^*$:
\begin{equation}
  \tilde{D}^{*} = 
  D^{*0} + \Rdsk e^{ i \left( \ddsk - \phi_3 \right) } \bar{D}^{*0}.
\end{equation}

We define $CP$ eigenstates of the neutral $D^*$ system, 
following the phase convention 
$CP (D^{*0}) = \bar{D}^{*0}$, 
$CP (\bar{D}^{*0}) = D^{*0}$~\cite{cp_phase}, 
so that
\begin{equation}
  D^*_+ = \frac{D^{*0} + \bar{D}^{*0}}{\sqrt{2}}, \hspace{8mm}
  D^*_- = \frac{D^{*0} - \bar{D}^{*0}}{\sqrt{2}},
\end{equation}
and
\begin{equation}
  D^{*0}       = \frac{D^*_+ + D^*_-}{\sqrt{2}},  \hspace{8mm}
  \bar{D}^{*0} = \frac{D^*_+ - D^*_-}{\sqrt{2}},
\end{equation}
and use similar definitions in the neutral $D$ system.
Thus
\begin{equation}
  \tilde{D}^* = 
  \frac{D^*_+ + D^*_-}{\sqrt{2}} +
  \Rdsk e^{ i \left( \ddsk - \phi_3 \right) } \frac{D^*_+ - D^*_-}{\sqrt{2}}.
\end{equation}

We now consider decays of the $D^*$ $CP$ eigenstates to $D\pi^0$ and $D\gamma$.
Using $\eta_X$ to denote the $CP$ eigenvalue of $X$,
we find that in the former case, $\eta_{D^*} = \eta_{D} \times \eta_{\pi^0} \times (-1)^l$,
where the angular momentum between the $D$ and $\pi^0$ mesons
is required to take the value $l = 1$ by conservation of angular momentum.
Thus $\eta_{D^*} = \eta_{D}$, and $D^*_\pm \to D_\pm \pi^0$.
In the latter case, we have $\eta_{D^*} = \eta_{D} \times \eta_{\gamma} \times (-1)^l$;
in this case we need to consider conservation of parity 
to find again that $l = 1$.
Thus $\eta_{D^*} = -1 \times \eta_{D}$, and so $D^*_\pm \to D_\mp \gamma$.

Next we consider the neutral $D$ meson produced
in the decay $B^- \to \tilde{D}^* K^-$, $\tilde{D}^* \to \tilde{D} \pi^0$,
\begin{eqnarray}
  \tilde{D} & = & 
  \frac{D_+ + D_-}{\sqrt{2}} + 
  \Rdsk e^{ i \left( \ddsk - \phi_3 \right) } \frac{D_+ - D_-}{\sqrt{2}} \\
  & = & 
  D^{0} + \Rdsk e^{ i \left( \ddsk - \phi_3 \right) } \bar{D}^{0},
\end{eqnarray}
whereas that produced in the decay $\tilde{D}^* \to \tilde{D} \gamma$ 
is given by
\begin{eqnarray}
  \tilde{D} & = & 
  \frac{D_- + D_+}{\sqrt{2}} +
  \Rdsk e^{ i \left( \ddsk - \phi_3 \right) } \frac{D_- - D_+}{\sqrt{2}} \\
  & = & 
  D^{0} - \Rdsk e^{ i \left( \ddsk - \phi_3 \right) } \bar{D}^{0} \\
  & = & 
  D^{0} + \Rdsk e^{ i \left( \ddsk + \pi - \phi_3 \right) } \bar{D}^{0}. 
\end{eqnarray}

Hence there is an effective strong phase shift of $\pi$ 
between the two cases~\cite{c_footnote}.

One should also consider the effect of other
nontrivial strong phases which may be produced in $D^*$ decay.
Although these may exist, 
we are interested only in the strong phase difference 
between $B^- \to D^{*0} K^-$ and $B^- \to \bar{D}^{*0} K^-$ amplitudes;
any such additional phase will be the same for 
both $D^{*0}$ and $\bar{D}^{*0}$ decays,
and thus cancel when the difference is considered,
due to $CP$ invariance of the $D^*$ decay amplitudes.

We now consider how this realization affects the various approaches
used for $\phi_3$ extraction.
In the GLW technique, the $D^*$ meson is reconstructed in $CP$ eigenstates.
If the $CP$ eigenvalues are properly taken into account, 
there is no advantage, except for statistical gain,
in using both $D^* \to D\pi^0$ and $D^* \to D\gamma$ decays.
However, experimentalists should take care to account for
cross-feed between these modes, which can pollute the $CP$ content.

In the Dalitz analysis technique, 
the strong phase shift of $\pi$ appears between the two $D^*$ decay modes,
and this should be taken into account in the analysis.
This method also allows a straightforward cross-check of our conclusion;
the strong phase can be measured from independent fits to samples 
with $D^* \to D \pi^0$ and $D^* \to D\gamma$, 
and the two cases should be found to differ by the factor of $\pi$.
The BaBar collaboration has recently released preliminary results
of such an analysis in which both $D^*$ decay modes are 
utilized~\cite{babar_dalitz}.
The treatment of the strong phases is not entirely clear;
in case the strong phase shift of $\pi$ is not taken into account the 
most likely effect is a reduction in the sensitivity to $\phi_3$.

It is for the ADS technique that the impact is most significant.
In this approach, it is convenient to measure the ratios of rates
for $B$ decays to suppressed and favoured final states, ${\cal R}$
so that the constant of proportionality drops out of Eq.~\ref{eq:gen_def}.
If we consider such ratios separately for $B^-$ and $B^+$, 
for both cases $D^* \to D\pi^0$ and $D^* \to D\gamma$, we see
\begin{eqnarray}
  {\cal R}_\mp \left( D^* \to D\pi^0 \right) & = &
  \Rdbsk^2 + \Rdbs^2 + 
  2 \Rdbsk \Rdbs \cos \left( \ddbsk + \ddbs \mp \phi_3 \right), \\
  {\cal R}_\mp \left( D^* \to D\gamma \right) & = &
  \Rdbsk^2 + \Rdbs^2 -
  2 \Rdbsk \Rdbs \cos \left( \ddbsk + \ddbs \mp \phi_3 \right).
\end{eqnarray}

\noindent
Note that here, in contrast to Eq.~\ref{eq:gen_def},
the strong phase difference of the $D^*$ decay is explicitly taken into account
so that $\delta_D$ is the strong phase difference of the $D$ decay amplitudes.

Since there are four independent equations,
which contain only three unknowns
(the value of $\Rdbs$ is known from $D$ decay), 
$\phi_3$ can be extracted, up to the usual four-fold ambiguity.
Compare this to the standard ADS analysis;
for any particular $B$ decay one finds only two independent equations,
which cannot be solved for three unknowns.
The usual resolution~\cite{hep-ph/0312100} is to combine decay modes, 
such as $B^\mp \to DK^\mp$ and $B^\mp \to D^*K^\mp$, 
but for each new mode two extra unknowns ($\Rdbsk$, $\ddbsk$) are added.

The charge averaged rates are given by
\begin{eqnarray}
  \label{eq:ads_dstar_dpi0}
  {\cal R} \left( D^* \to D\pi^0 \right) & = &
  \Rdbsk^2 + \Rdbs^2 + 
  2 \Rdbsk \Rdbs \cos \left( \ddbsk + \ddbs \right) \cos \left( \phi_3 \right), \\
  \label{eq:ads_dstar_dgamma}
  {\cal R} \left( D^* \to D\gamma \right) & = &
  \Rdbsk^2 + \Rdbs^2 - 
  2 \Rdbsk \Rdbs \cos \left( \ddbsk + \ddbs \right) \cos \left( \phi_3 \right),
\end{eqnarray}
so that 
\begin{equation}
  \label{eq:ads_dstar_Rav}
  \left(
    {\cal R} \left( D^* \to D\pi^0 \right) +
    {\cal R} \left( D^* \to D\gamma \right)
  \right)/2 = \Rdbsk^2 + \Rdbs^2,
\end{equation}
does not depend on any phases.
Hence the value of $\Rdbsk$ in $B^- \to D^*K^-$ decays can be 
straightforwardly obtained.

Recently, the BaBar collaboration has performed an ADS analysis 
using $B^\mp \to D^* K^\mp$,
using the subsequent doubly Cabibbo-suppressed decay $D \to K\pi$~\cite{babar_ads}.
In the absence of any significant signals, 
they combine the likelihoods for 
${\cal R} \left( D^* \to D\pi^0 \right)$ and ${\cal R} \left( D^* \to D\gamma \right)$,
and obtain an upper limit on the quantity on the left-hand side of
Eq.~\ref{eq:ads_dstar_Rav} of 0.021 at 90\% confidence level (C.L.).
This is then converted into an upper limit of $\Rdbsk < 0.21$ (90\% C.L.).
(For the suppressed $K\pi$ final state, 
the ratio of $D$ decay amplitudes is known to be 
$r_D = 0.060 \pm 0.003$~\cite{pdg}.)
Since the strong phase shift of $\pi$ is not taken into account,
this limit is calculated using Eq.~\ref{eq:ads_dstar_dpi0},
and allowing for any possible values of $\ddbsk + \ddbs$ and $\phi_3$.
However, if the upper limit is calculated using Eq.~\ref{eq:ads_dstar_Rav},
a more stringent value of $\Rdbsk < 0.134$ can be obtained.

In conclusion, we have shown that in decays of the type $B^- \to D^* K^-$,
there is an effective strong phase difference of $\pi$ between 
the cases that the $D^*$ is reconstructed as $D\pi^0$ and $D\gamma$.
We have examined the consequences for several approaches to extract $\phi_3$, 
and find a significant benefit for the ADS technique.
In future this enhanced ADS method may be used to measure $\phi_3$
using $B^\mp \to D^* K^\mp$ decays alone.

We are grateful to V.~Chernyak, A.~Vainshtein, M.~Voloshin and H.~Yamamoto
for useful discussions.
T.~G. is supported by the Japan Society for the Promotion of Science.
This work was carried out during a visit to Japan by A.~B., supported by KEK.


\begin{thebibliography}{99}

\bibitem{pdg_review}
  For a review of $CP$ violation phenomenology, see 
  D.~Kirkby and Y.~Nir in~\cite{pdg}.

\bibitem{pdg}
  S.~Eidelman {\it et al.}, Phys. Lett. B {\bf 592}, 1 (2004).

\bibitem{glw}
  M.~Gronau and D.~London, Phys. Lett. B {\bf 253}, 483 (1991),
  M.~Gronau and D.~Wyler, Phys. Lett. B {\bf 265}, 172 (1991).

\bibitem{ads}
  D.~Atwood, I.~Dunietz, and A.~Soni, Phys. Rev. Lett. {\bf 78}, 3257 (1997),
  Phys. Rev. D {\bf 63}, 036005 (2001).

\bibitem{dalitz}
  A.~Giri, Y.~Grossman, A.~Soffer and J.~Zupan,
  Phys. Rev. D {\bf 68}, 054018 (2003);
  A.~Poluektov {\it et al.} (Belle Collaboration), 
  {\tt hep-ex/0406067} (to appear in Phys. Rev. D).
  
\bibitem{cp_phase}
  The choice of $CP$ phase convention does not affect observable quantities.

\bibitem{c_footnote}
  This conclusion can also be obtained invoking only $C$ symmetry
  in $D^*$ decays.

\bibitem{babar_dalitz}
  B.~Aubert {\it et al.} (BaBar Collaboration),
  BaBar-CONF-04/043, {\tt hep-ex/0408088}.

\bibitem{hep-ph/0312100}
  D.~Atwood and A.~Soni, {\tt hep-ph/0312100}.

\bibitem{babar_ads}
  B.~Aubert {\it et al.} (BaBar Collaboration),
  BaBar-CONF-04/013, {\tt hep-ex/0408028}.

\end{thebibliography}
\end{document}